\documentclass[12pt,twoside]{article}
\usepackage[english]{babel}
\usepackage{graphicx,rotating,epsfig}
\usepackage{a4p}

\def\ra{\rightarrow}
\def\GeV  {\ensuremath{\mathrm{ Ge\kern -0.1em V } }}
\def\GeVc2{\ensuremath{\mathrm{ Ge\kern -0.1em V }\kern -0.2em /c^2 }}
\def\MeVc2{\ensuremath{\mathrm{ Me\kern -0.1em V }\kern -0.2em /c^2 }}
\newcommand{\MT}{\ensuremath{M_{\mathrm{t}}}}
\newcommand{\MTll}{\ensuremath{\MT^{\mathrm{di-l}}}}
\newcommand{\MTlj}{\ensuremath{\MT^{\mathrm{l+j}}}}
\newcommand{\MTjj}{\ensuremath{\MT^{\mathrm{all-j}}}}

\newcommand{\pb}{\ensuremath{\mathrm{pb}^{-1}}}
\newcommand{\fb}{\ensuremath{\mathrm{fb}^{-1}}}
\newcommand{\ttbar}{\ensuremath{t\overline{t}}}
\newcommand{\ljt}{\ensuremath{\ell\nu q q^{\prime} b \overline{b}}}
\newcommand{\had}{\ensuremath{q q^{\prime} b q q^{\prime} \overline{b}}}
\newcommand{\dil}{\ensuremath{\ell^{+}\nu b\ell^{-}\overline{\nu}\overline{b}}}
\newcommand{\ttljt}{\ensuremath{\ttbar\ra\ljt}}
\newcommand{\ttdil}{\ensuremath{\ttbar\ra\dil}}
\newcommand{\tthad}{\ensuremath{\ttbar\ra\had}}
\newcommand{\RunI}{\hbox{Run-I}}
\newcommand{\RunII}{\hbox{Run-II}}

\begin{document}

\begin{center}
  {\LARGE FERMI NATIONAL ACCELERATOR LABORATORY}
\end{center}

\begin{flushright}
       FERMILAB-TM-2403-E \\  
       TEVEWWG/top 2008/01 \\
       CDF Note 9225 \\
       D\O\ Note 5626 \\
       \vspace*{0.05in}
       6th March 2008 \\
\end{flushright}

\vskip 1cm

\begin{center}
  {\LARGE\bf 
    Combination of CDF and D\O\ Results \\
    on the Mass of the Top Quark\\
  }
  \vfill
  {\Large
    The Tevatron Electroweak Working Group\footnote{The Tevatron Electroweak 
    Working Group can be contacted at tev-ewwg@fnal.gov.\\  
    \hspace*{0.20in} More information can
    be found at {\tt http://tevewwg.fnal.gov}.} \\
    for the CDF and D\O\ Collaborations\\
  }
\end{center}
\vfill
\begin{abstract}
\noindent
  We summarize the top-quark mass measurements from the CDF and
  D\O\ experiments at Fermilab.  We combine published
  \RunI\ (1992-1996) measurements with the most recent preliminary
  \RunII\ (2001-present) measurements using up to $2.1~\fb$ of data.
  Taking correlated uncertainties properly into account the resulting
  preliminary world average mass of the top quark is
  $\MT=172.6\pm0.8\mathrm{(stat)}\pm1.1\mathrm{(syst)}~\GeVc2$,
  assuming Gaussian systematic uncertainties. Adding in quadrature
  yields a total uncertainty of $1.4~\GeVc2$, corresponding to a
  relative precision of 0.8\% on the top-quark mass.
\end{abstract}

\vfill



\section{Introduction}
\label{sec:intro}

The experiments CDF and D\O, taking data at the Tevatron
proton-antiproton collider located at the Fermi National Accelerator
Laboratory, have made several direct experimental measurements of the
top-quark pole mass, \MT.  The pioneering measurements were based on about
$100~\pb$ of \RunI\ (1992-1996) data~\cite{Mtop1-CDF-di-l-PRLa, 
  Mtop1-CDF-di-l-PRLb,
  Mtop1-CDF-di-l-PRLb-E, Mtop1-D0-di-l-PRL, Mtop1-D0-di-l-PRD,
  Mtop1-CDF-l+j-PRL, Mtop1-CDF-l+j-PRD, Mtop1-D0-l+j-old-PRL,
  Mtop1-D0-l+j-old-PRD, Mtop1-D0-l+j-new1, Mtop1-CDF-all-j-PRL,
  Mtop1-D0-all-j-PRL} 
and include results from the \tthad\ (all-j), the \ttljt\ (l+j), and the 
\ttdil\ (di-l) decay channels\footnote{Here $\ell=e$ or $\mu$.  Decay 
channels with explicit tau lepton identification are presently under 
study and are not yet used for measurements of the top-quark mass.}.   
The \RunII\ measurements summarized here are the most recent results in the 
l+j, di-l,  and all-j channels using $1.9-2.1~\fb$ of data and improved 
analysis techniques~\cite{
Mtop2-CDF-lxy-new,
Mtop2-CDF-di-l-new,
Mtop2-CDF-l+j-new,
Mtop2-CDF-all-j-new, 
Mtop2-D0-l+ja-new,
Mtop2-D0-l+jb-new,
Mtop2-D0-di-l-nwt,
Mtop2-D0-di-l-mwt,
Mtop2-D0-di-l-new}.  
\vspace*{0.10in}

This note reports the world average top-quark mass obtained by
combining five published
\RunI\ measurements~\cite{Mtop1-CDF-di-l-PRLb, Mtop1-CDF-di-l-PRLb-E,
  Mtop1-D0-di-l-PRD, Mtop1-CDF-l+j-PRD, Mtop1-D0-l+j-new1,
  Mtop1-CDF-all-j-PRL} with one published \RunII\ CDF
result~\cite{Mtop2-CDF-lxy-new}, three preliminary \RunII\ CDF
results~\cite{Mtop2-CDF-di-l-new, Mtop2-CDF-l+j-new, Mtop2-CDF-all-j-new}
and three preliminary
\RunII\ D\O\ results~\cite{Mtop2-D0-l+ja-new, Mtop2-D0-l+jb-new,
  Mtop2-D0-di-l-new}.  The combination takes into account the
statistical and systematic uncertainties and their correlations using
the method of references~\cite{Lyons:1988, Valassi:2003} and
supersedes previous
combinations~\cite{Mtop1-tevewwg04,Mtop-tevewwgSum05,
  Mtop-tevewwgWin06,Mtop-tevewwgSum06, Mtop-tevewwgWin07}.
\vspace*{0.10in}

The input measurements and error categories used in the combination are 
detailed in Section~\ref{sec:inputs} and~\ref{sec:errors}, respectively. 
The correlations used in the combination are discussed in 
Section~\ref{sec:corltns} and the resulting world average top-quark mass 
is given in Section~\ref{sec:results}.  A summary and outlook are presented
in Section~\ref{sec:summary}.
 
\section{Input Measurements}
\label{sec:inputs}

For this combination twelve measurements of \MT\ are used, five
published \RunI\ results, and one published plus six preliminary
\RunII\ results, all reported in Table~\ref{tab:inputs}.  In general,
the \RunI\ measurements all have relatively large statistical
uncertainties and their systematic uncertainty is dominated by the
total jet energy scale (JES) uncertainty.  In \RunII\ both CDF and
D\O\ take advantage of the larger \ttbar\ samples available and employ
new analysis techniques to reduce both these uncertainties.  In
particular the JES is constrained using an in-situ calibration based
on the invariant mass of $W\ra qq^{\prime}$ decays in the l+j and
all-j channels.  The \RunII\ D\O\ analysis in the l+j channel
constrains the response of light-quark jets using the in-situ $W\ra
qq^{\prime}$ decays.  Residual JES uncertainties associated with
$\eta-$ and $p_{T}$-dependencies as well as uncertainties specific to
the response of $b$-jets are treated separately.  Similarly, the
\RunII\ CDF analysis in the l+j and all-j channels also constrain the
JES using the in-situ $W\ra qq^{\prime}$ decays.  Small residual JES
uncertainties arising from $\eta-$ and $p_{T}$-dependencies and the
modeling of $b$-jets are included in separate error categories.  The
\RunII\ CDF di-l measurement uses a JES determined from external
calibration samples.  Some parts of the associated uncertainty are
correlated with the \RunI\ JES uncertainty as noted below.
\vspace*{0.10in}

\begin{table}[t]
\begin{center}
\renewcommand{\arraystretch}{1.30}
\begin{tabular}{|l||rrr|rr||rrrr|rrr|}
\hline       
       & \multicolumn{5}{|c||}{{\RunI} published} & \multicolumn{7}{|c|}{{\RunII} preliminary} \\ \cline{2-13}
       & \multicolumn{3}{|c|}{ CDF } & \multicolumn{2}{|c||}{ D\O\ }
       & \multicolumn{4}{|c|}{ CDF } & \multicolumn{3}{|c|}{ D\O\ } \\
       & all-j & l+j   & di-l  & l+j   & di-l  & l+j   & di-l  & all-j & lxy   & l+j/a & l+j/b & di-l \\
\hline                         
Result & 186.0 & 176.1 & 167.4 & 180.1 & 168.4 & 172.7 & 171.2 & 177.0 & 180.7 & 170.5 & 173.0 & 173.7 \\
\hline                         
\hline                         
iJES   &   0.0 &   0.0 &   0.0 &   0.0 &   0.0 &   1.3 &   0.0 &   1.8 &   0.0 &   0.0 &   0.0 &   0.0 \\
aJES   &   0.0 &   0.0 &   0.0 &   0.0 &   0.0 &   0.0 &   0.0 &   0.0 &   0.0 &   0.7 &   0.8 &   1.9 \\
bJES   &   0.6 &   0.6 &   0.8 &   0.7 &   0.7 &   0.4 &   0.1 &   0.1 &   0.0 &   0.2 &   0.1 &   0.9 \\
cJES   &   3.0 &   2.7 &   2.6 &   2.0 &   2.0 &   0.5 &   1.7 &   0.6 &   0.0 &   0.0 &   0.0 &   2.1 \\
dJES   &   0.3 &   0.7 &   0.6 &   0.0 &   0.0 &   0.1 &   0.1 &   0.1 &   0.0 &   1.7 &   1.4 &   0.9 \\
rJES   &   4.0 &   3.4 &   2.7 &   2.5 &   1.1 &   0.2 &   1.8 &   0.5 &   0.3 &   0.0 &   0.0 &   0.0 \\
Signal &   1.8 &   2.6 &   2.8 &   1.1 &   1.8 &   0.6 &   0.7 &   0.6 &   1.4 &   1.0 &   0.5 &   0.8 \\
BG     &   1.7 &   1.3 &   0.3 &   1.0 &   1.1 &   0.6 &   0.4 &   1.0 &   7.2 &   0.5 &   0.4 &   0.6 \\
Fit    &   0.6 &   0.0 &   0.7 &   0.6 &   1.1 &   0.2 &   0.6 &   0.6 &   4.2 &   0.1 &   0.2 &   0.9 \\
MC     &   0.8 &   0.1 &   0.6 &   0.0 &   0.0 &   0.4 &   0.7 &   0.3 &   0.7 &   0.0 &   0.0 &   0.2 \\
UN/MI  &   0.0 &   0.0 &   0.0 &   1.3 &   1.3 &   0.0 &   0.0 &   0.0 &   0.0 &   0.0 &   0.0 &   0.0 \\
\hline                         
Syst.  &   5.7 &   5.3 &   4.9 &   3.9 &   3.6 &   1.7 &   2.8 &   2.4 &   8.5 &   2.2 &   1.7 &   3.4 \\
Stat.  &  10.0 &   5.1 &  10.3 &   3.6 &  12.3 &   1.2 &   2.7 &   3.3 &  14.5 &   1.9 &   1.3 &   5.4 \\
\hline                         
\hline                         
Total  &  11.5 &   7.3 &  11.4 &   5.3 &  12.8 &   2.1 &   3.9 &   4.1 &  16.8 &   2.9 &   2.2 &   6.4 \\ 
\hline
\end{tabular}
\end{center}
\caption[Input measurements]{Summary of the measurements used to determine the
  world average $\MT$.  All numbers are in $\GeVc2$.  The error categories and 
  their correlations are described in the text.  The total systematic uncertainty 
  and the total uncertainty are obtained by adding the relevant contributions 
  in quadrature.}
\label{tab:inputs}
\end{table}

In previous combinations the \RunII\ CDF l+j analysis used the JES
determined from the external calibration as an additional Gaussian
constraint.  This required us to treat that measurement as two
separate inputs in the combination in order to accurately account for
all the JES correlations.  This Gaussian constraint is not used in the
present analysis as it does not significantly improve the sensitivity.
Thus we can treat this measurement as a single input in the same
manner as all the other measurements.
\vspace*{0.10in}

As discussed in the previous combination, a new analysis technique
from CDF is included (lxy).  This measurement uses the mean
decay-length from B-tagged jets to determine the top-quark mass.
While the statistical sensitivity is not nearly as good as the more
traditional methods, this technique has the advantage that since it
uses only tracking information, it is almost entirely independent of
JES uncertainties.  As the statitistics of this sample continue to
grow, this method could offer a nice cross-check of the top-quark mass
that's largely independent of the dominant JES systematic uncertainty
which plagues the other measurements.  The statistical correlation
between this measurement and the \RunII\ CDF l+j measurement is
determined using Monte Carlo signal-plus-background psuedo-experiments
which correctly account for the sample overlap and is found to be
consistent with zero (to within $<1\%$) independent of the assumed
top-quark mass.
\vspace*{0.10in}

The two D\O\ Run-II lepton+jets results~\cite{Mtop2-D0-l+ja-new,
  Mtop2-D0-l+jb-new} are derived from Run-IIa and Run-IIb datasets,
respectively, and are labelled as such.  The D\O\ Run-II dilepton
result~\cite{Mtop2-D0-di-l-new} is itself a combination of two results
using different techniques but the same di-lepton data
set~\cite{Mtop2-D0-di-l-nwt,Mtop2-D0-di-l-mwt}.
\vspace*{0.10in}

Table~\ref{tab:inputs} also lists the uncertainties of the results,
sub-divided into the categories described in the next Section.  The
correlations between the inputs are described in
Section~\ref{sec:corltns}.


\section{Error Categories}
\label{sec:errors}

We employ the same error categories as used for the previous world
average~\cite{Mtop-tevewwgWin07}.  They include a detailed
breakdown of the various sources of uncertainty and aim to
lump together sources of systematic uncertainty that share the same or
similar origin.  For example, the ``Signal'' category discussed below
includes the uncertainties from ISR, FSR, and PDF - all of which affect
the modeling of the \ttbar\ signal.  Additional categories are included 
in order to accommodate specific types of correlations.  For example,
the jet energy scale (JES) uncertainty is sub-divided into several
components in order to more accurately accommodate our best estimate of
the relevant correlations.  Each error category is discussed below.
\vspace*{0.10in}

\begin{description}
  \item[Statistical:] The statistical uncertainty associated with the
    \MT\ determination.
  \item[iJES:] That part of the JES uncertainty which originates from 
    in-situ calibration procedures and is uncorrelated among the
    measurements.  In the combination reported here it corresponds to  
    the statistical uncertainty associated with the JES determination 
    using the $W\ra qq^{\prime}$ invariant mass in the CDF \RunII\ 
    l+j and all-h measurements.  Residual JES uncertainties, which arise 
    from effects 
    not considered in the in-situ calibration, are included in other 
    categories.
  \item[aJES:] That part of the JES uncertainty which originates from
    differences in detector $e/h$ response between $b$-jets and light-quark
    jets.  It is specific to the D\O\ \RunII\ measurements and is
    taken to be uncorrelated with the D\O\ \RunI\ and CDF measurements.
  \item[bJES:] That part of the JES uncertainty which originates from
    uncertainties specific to the modeling of $b$-jets and which is correlated
    across all measurements.  For both CDF and D\O\ this includes uncertainties 
    arising from 
    variations in the semi-leptonic branching fraction, $b$-fragmentation 
    modeling, and differences in the color flow between $b$-jets and light-quark
    jets.  These were determined from \RunII\ studies but back-propagated
    to the \RunI\ measurements, whose rJES uncertainties (see below) were 
    then corrected in order to keep the total JES uncertainty constant.
  \item[cJES:] That part of the JES uncertainty which originates from
    modeling uncertainties correlated across all measurements.  Specifically
    it includes the modeling uncertainties associated with light-quark 
    fragmentation and out-of-cone corrections.
  \item[dJES:] That part of the JES uncertainty which originates from
    limitations in the calibration data samples used and which is
    correlated between measurements within the same data-taking
    period, such as Run~I, Run~IIa or Run~IIb, but not between
    experiments.  For CDF this corresponds to uncertainties associated
    with the $\eta$-dependent JES corrections which are estimated
    using di-jet data events.  For D\O\ \RunII\ this corresponds to
    uncertainties associated with the light-quark response as
    determined using the $W\ra qq^{\prime}$ invariant mass in the l+j
    channel and propagated to the di-l channel.  The residual
    $\eta$-dependent and $p_{T}$-dependent uncertainties for the D\O\
    \RunII\ measurements are also included here since they are
    constrained using \RunII\ $\gamma+$jet data samples.
  \item[rJES:] The remaining part of the JES uncertainty which is 
    correlated between all measurements of the same experiment 
    independent of data-taking period, but is uncorrelated between
    experiments.  This is dominated by uncertainties in the calorimeter
    response to light-quark jets.  For CDF this also includes small 
    uncertainties associated with the multiple interaction and underlying 
    event corrections.
  \item[Signal:] The systematic uncertainty arising from uncertainties
    in the modeling of the \ttbar\ signal which is correlated across all
    measurements.  This includes uncertainties from variations in the ISR,
    FSR, and PDF descriptions used to generate the \ttbar\ Monte Carlo samples
    that calibrate each method.  It also includes small uncertainties 
    associated with biases associated with the identification of $b$-jets.
  \item[Background:]  The systematic uncertainty arising from uncertainties
    in modeling the dominant background sources and correlated across
    all measurements in the same channel.  These
    include uncertainties on the background composition and shape.  In
    particular uncertainties associated with the modeling of the QCD
    multi-jet background (all-j and l+j), uncertainties associated with the
    modeling of the Drell-Yan background (di-l), and uncertainties associated 
    with variations of the fragmentation scale used to model W+jets 
    background (all channels) are included.
  \item[Fit:] The systematic uncertainty arising from any source specific
    to a particular fit method, including the finite Monte Carlo statistics 
    available to calibrate each method.
  \item[Monte Carlo:] The systematic uncertainty associated with variations
    of the physics model used to calibrate the fit methods and correlated
    across all measurements.  For CDF it includes variations observed when 
    substituting PYTHIA~\cite{PYTHIA4,PYTHIA5,PYTHIA6} (Run~I and Run~II) 
    or ISAJET~\cite{ISAJET} (Run~I) for HERWIG~\cite{HERWIG5,HERWIG6} when 
    modeling the \ttbar\ signal.  Similar
    variations are included for the D\O\ \RunI\ measurements.  The D\O\ 
    \RunII\ measurements use ALPGEN~\cite{ALPGEN} to model the \ttbar\ signal and the
    variations considered are included in the Signal category above.
  \item[UN/MI:] This is specific to D\O\ and includes the uncertainty
    arising from uranium noise in the D\O\ calorimeter and from the
    multiple interaction corrections to the JES.  For D\O\ \RunI\ these
    uncertainties were sizable, while for \RunII\, owing to the shorter
    integration time and in-situ JES determination, these uncertainties
    are negligible.
\end{description}
These categories represent the current preliminary understanding of the
various sources of uncertainty and their correlations.  We expect these to 
evolve as we continue to probe each method's sensitivity to the various 
systematic sources with ever improving precision.  Variations in the assignment
of uncertainties to the error categories, in the back-propagation of the bJES
uncertainties to \RunI\ measurements, in the approximations made to
symmetrize the uncertainties used in the combination, and in the assumed 
magnitude of the correlations all negligibly effect ($\ll 0.1\GeVc2$) the 
combined \MT\ and total uncertainty.

\section{Correlations}
\label{sec:corltns}

The following correlations are used when making the combination:
\begin{itemize}
  \item The uncertainties in the Statistical, Fit, and iJES
    categories are taken to be uncorrelated among the measurements.
  \item The uncertainties in the aJES and dJES categories are taken
    to be 100\% correlated among all \RunI\ and all \RunII\ measurements 
    on the same experiment, but uncorrelated between Run~I and Run~II
    and uncorrelated between the experiments.
  \item The uncertainties in the rJES and UN/MI categories are taken
    to be 100\% correlated among all measurements on the same experiment.
  \item The uncertainties in the Background category are taken to be
    100\% correlated among all measurements in the same channel.
  \item The uncertainties in the bJES, cJES, Signal, and Generator
    categories are taken to be 100\% correlated among all measurements.
\end{itemize}
Using the inputs from Table~\ref{tab:inputs} and the correlations specified
here, the resulting matrix of total correlation co-efficients is given in
Table~\ref{tab:coeff}.

\begin{table}[t]
\begin{center}
\renewcommand{\arraystretch}{1.30}
\begin{tabular}{|ll||rrr|rr||rrrr|rrr|}
\hline       
   &   & \multicolumn{5}{|c||}{{\RunI} published} & \multicolumn{7}{|c|}{{\RunII} preliminary} \\ \cline{3-14}
   &   & \multicolumn{3}{|c|}{ CDF } & \multicolumn{2}{|c||}{ D\O\ }
       & \multicolumn{4}{|c|}{ CDF } & \multicolumn{3}{|c|}{ D\O\ } \\
   &            & l+j & di-l & all-j &   l+j &  di-l & l+j   & di-l  & all-j & lxy & l+j/a & l+j/b & di-l \\
\hline
\hline
CDF-I & l+j     & 1.00&      &       &       &       &       &       &      &      &      &      & \\
CDF-I & di-l    & 0.29&  1.00&       &       &       &       &       &      &      &      &      & \\
CDF-I & all-j   & 0.32&  0.19&   1.00&       &       &       &       &      &      &      &      & \\
\hline
D\O-I & l+j     & 0.26&  0.15&   0.14&   1.00&       &       &       &      &      &      &      & \\
D\O-I & di-l    & 0.11&  0.08&   0.07&   0.16&   1.00&       &       &      &      &      &      & \\
\hline
\hline
CDF-II & l+j    & 0.30&  0.17&   0.16&   0.22&   0.09&   1.00&       &      &      &      &      & \\
CDF-II & di-l   & 0.45&  0.27&   0.33&   0.21&   0.11&   0.24&   1.00&      &      &      &      & \\
CDF-II & all-j  & 0.17&  0.11&   0.15&   0.09&   0.05&   0.11&   0.17&  1.00&      &      &      & \\
CDF-II & lxy    & 0.11&  0.03&   0.02&   0.10&   0.01&   0.16&   0.03&  0.02&  1.00&      &      & \\
\hline
D\O-II & l+j/a  & 0.16&  0.09&   0.06&   0.11&   0.05&   0.16&   0.07&  0.06&  0.10&  1.00&      & \\
D\O-II & l+j/b  & 0.11&  0.06&   0.04&   0.09&   0.03&   0.12&   0.04&  0.04&  0.09&  0.20& 1.00 & \\
D\O-II & di-l   & 0.18&  0.12&   0.11&   0.17&   0.08&   0.14&   0.19&  0.07&  0.01&  0.21& 0.14 & 1.00\\
\hline
\end{tabular}
\end{center}
\caption[Global correlations between input measurements]{The resulting
  matrix of total correlation coefficients used to determined the
  world average top quark mass.}
\label{tab:coeff}
\end{table}

The measurements are combined using a program implementing a numerical
$\chi^2$ minimization as well as the analytic BLUE
method~\cite{Lyons:1988, Valassi:2003}. The two methods used are
mathematically equivalent, and are also equivalent to the method used
in an older combination~\cite{TM-2084}, and give identical results for
the combination. In addition, the BLUE method yields the decomposition
of the error on the average in terms of the error categories specified
for the input measurements~\cite{Valassi:2003}.

\section{Results}
\label{sec:results}

\begin{figure}[p]
\begin{center}
\includegraphics[width=0.8\textwidth]{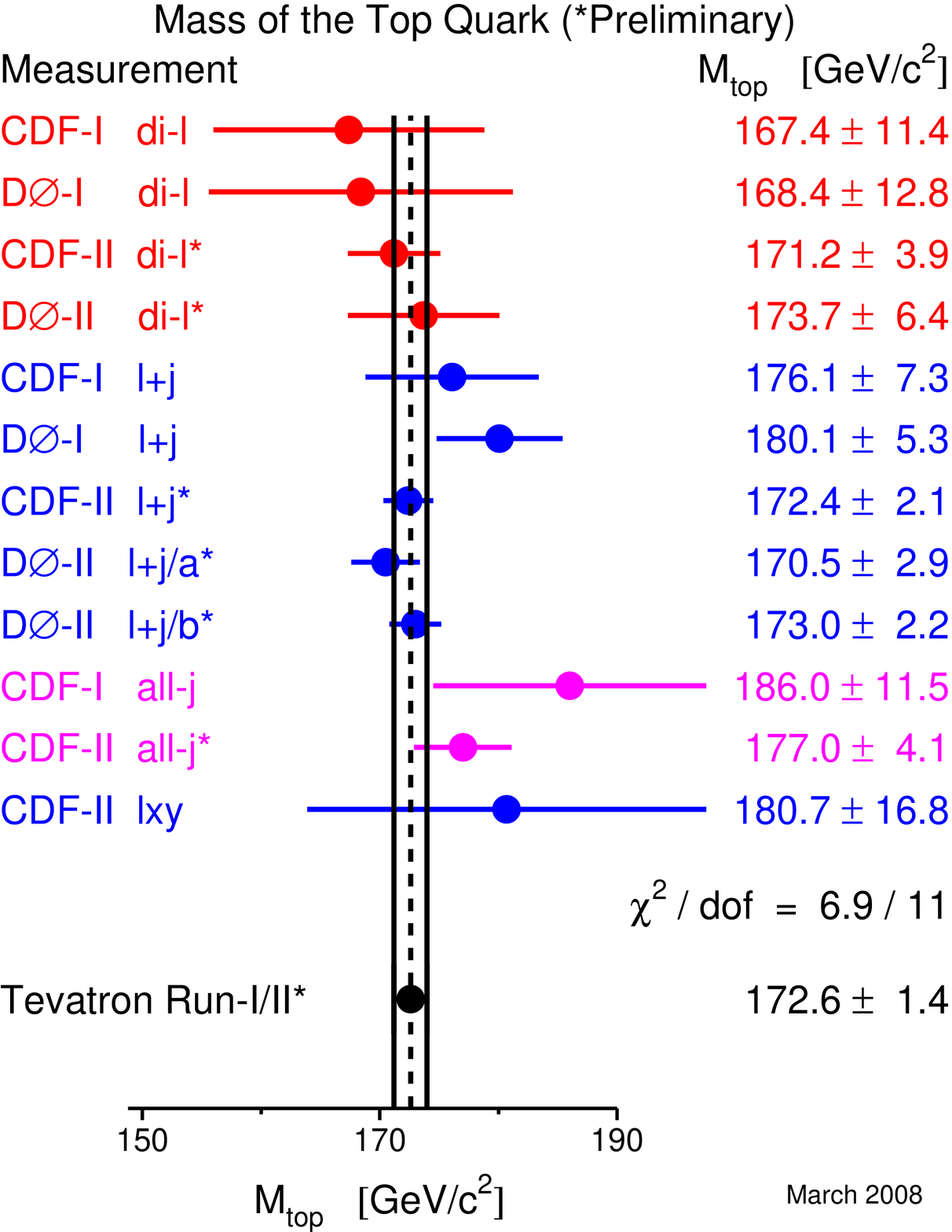}
\end{center}
\caption[Summary plot for the world average top-quark mass]
  {A summary of the input measurements and resulting world average
   mass of the top quark.}
\label{fig:summary} 
\end{figure}

The combined value for the top-quark mass is:
\begin{eqnarray}
  \MT & = & 172.6 \pm 1.4~\GeVc2\,,
\end{eqnarray}
with a $\chi^2$ of 6.9 for 11 degrees of freedom, which corresponds to
a probability of 81\% indicating good agreement among all the input
measurements.  The total uncertainty can be sub-divided into the 
contributions from the various error categories as: Statistical ($\pm0.8$),
total JES ($\pm0.9$), Signal ($\pm0.5$), Background ($\pm0.4$), Fit
($\pm0.1$), Monte Carlo ($\pm0.2$), and UN/MI ($\pm0.02$), for a total
Systematic ($\pm1.1$), where all numbers are in units of \GeVc2.
The pull and weight for each of the inputs are listed in Table~\ref{tab:stat}.
The input measurements and the resulting world average mass of the top 
quark are summarized in Figure~\ref{fig:summary}.
\vspace*{0.10in}

The weights of many of the \RunI\ measurements are negative. 
In general, this situation can occur if the correlation between two measurements
is larger than the ratio of their total uncertainties. This is indeed the case
here.  In these instances the less precise measurement 
will usually acquire a negative weight.  While a weight of zero means that a
particular input is effectively ignored in the combination, a negative weight 
means that it affects the resulting central value and helps reduce the total
uncertainty. See reference~\cite{Lyons:1988} for further discussion of 
negative weights.

\begin{table}[t]
\begin{center}
\renewcommand{\arraystretch}{1.30}
\begin{tabular}{|l||rrr|rr||rrrr|rrr|}
\hline       
       & \multicolumn{5}{|c||}{{\RunI} published} & \multicolumn{7}{|c|}{{\RunII} preliminary} \\ \cline{2-13}
       & \multicolumn{3}{|c|}{ CDF } & \multicolumn{2}{|c||}{ D\O\ }
       & \multicolumn{4}{|c|}{ CDF } & \multicolumn{3}{|c|}{ D\O\ } \\
       & l+j     & di-l    & all-j   & l+j     & di-l    & l+j     & di-l    & all-j   & lxy 
       & l+j/a   & l+j/b   & di-l\\
\hline
\hline
Pull   & $+0.5$  & $-0.5$  & $+1.2$  & $+1.5$  & $-0.3$  & $+0.1$  & $-0.4$  & $+1.2$  & $+0.5$    
       & $-0.8$  & $+0.2$  & $+0.2$ \\
Weight [\%]
       & $- 4.2$ & $- 0.7$ & $- 0.6$ & $+ 1.8$ & $+ 0.2$ & $+35.8$ & $+ 9.7$ & $+8.8$ & $- 0.7$        
       & $+15.2$ & $+35.2$ & $-0.6$  \\
\hline
\end{tabular}
\end{center}
\caption[Pull and weight of each measurement]{The pull and weight for each of the
  inputs used to determine the world average mass of the top quark.  See 
  Reference~\cite{Lyons:1988} for a discussion of negative weights.}
\label{tab:stat} 
\end{table} 

Although the $\chi^2$ from the combination of all measurements indicates
that there is good agreement among them, and no input has an anomalously
large pull, it is still interesting to also fit for the top-quark mass
in the all-j, l+j, and di-l channels separately.  We use the same methodology,
inputs, error categories, and correlations as described above, but fit for
the three physical observables, \MTjj, \MTlj, and \MTll.
The results of this combination are shown in Table~\ref{tab:three_observables}
and have $\chi^2$ of 4.2 for 9 degrees of freedom, which corresponds to a
probability of 90\%.
These results differ from a naive combination, where
only the measurements in a given channel contribute to the \MT\ 
determination in that channel, since the combination here fully accounts
for all correlations, including those which cross-correlate the different
channels. Using the results of 
Table~\ref{tab:three_observables} we calculate the chi-squared consistency
between any two channels, including all correlations, as 
$\chi^{2}(dil-lj)=0.8$, $\chi^{2}(lj-allj)=1.5$, and 
$\chi^{2}(allj-dil)=2.7$.  These correspond to 
chi-squared probabilities of 39\%, 23\%, and 10\%, respectively, and indicate 
that the determinations of \MT\ from the three channels are consistent with 
one another.

\begin{table}[t]
\begin{center}
\renewcommand{\arraystretch}{1.30}
\begin{tabular}{|l||c|rrr|}
\hline
Parameter & Value (\GeVc2) & \multicolumn{3}{|c|}{Correlations} \\
\hline
\hline
$\MTjj$ & $177.3\pm 3.9$ & 1.00 &      &      \\
$\MTlj$ & $172.4\pm 1.5$ & 0.12 & 1.00 &      \\
$\MTll$ & $169.8\pm 3.1$ & 0.18 & 0.26 & 1.00 \\
\hline
\end{tabular}
\end{center}
\caption[Mtop in each channel]{Summary of the combination of the nine
measurements by CDF and D\O\ in terms of three physical quantities,
the mass of the top quark in the all-jets, lepton+jets, and di-lepton channel. }
\label{tab:three_observables}
\end{table}

\section{Summary}
\label{sec:summary}

A preliminary combination of measurements of the mass of the top quark
from the Tevatron experiments CDF and D\O\ is presented.  The
combination includes five published {\RunI} measurements and one
published plus six preliminary {\RunII} measurements.  Taking into
account the statistical and systematic uncertainties and their
correlations, the preliminary world-average result is: $\MT= 172.6 \pm
1.4~\GeVc2$, where the total uncertainty is obtained assuming Gaussian
systematic uncertainties and adding them plus the statistical
uncertainty in quadrature.  While the central value is somewhat higher
than our 2007 average, the averages are compatible as appreciably more
luminosity and refined analysis techniques are now used.
\vspace*{0.10in}

The mass of the top quark is now known with a relative precision of
0.8\%, limited by the systematic uncertainties, which are dominated by
the jet energy scale uncertainty.  This systematic is expected to
improve as larger data sets are collected since new analysis
techniques constrain the jet energy scale using in-situ $W\ra
qq^{\prime}$ decays. It can be reasonably expected that with the full
\RunII\ data set the top-quark mass will be known to much better than
0.8\%.  To reach this level of precision further work is required to
determine more accurately the various correlations present, and to
understand more precisely the $b$-jet modeling, Signal, and Background
uncertainties which may limit the sensitivity at larger data sets.
Limitations of the Monte Carlo generators used to calibrate each fit
method may also become important as the precision reaches the
$\sim1~\GeVc2$ level and will warrant further study in the near
future.

\clearpage

\bibliographystyle{tevewwg}
\bibliography{run2mtop}

\end{document}